\documentclass[aps, superscriptaddress, showpacs, floatfix, twocolumn]{revtex4-1}
\usepackage{amsmath}
\usepackage{hyperref}
\usepackage{graphicx}
\begin{document}
\title{Photoinduced phase transitions in narrow-gap Mott\\ insulators: the case of VO$_2$}
\author{Zhuoran He}
\affiliation{Department of Physics, Columbia University, New York, New York 10027, USA}
\author{Andrew J. Millis}
\affiliation{Department of Physics, Columbia University, New York, New York 10027, USA}

\begin{abstract}
\vspace*{0\baselineskip}
We study the nonequilibrium dynamics of photoexcited electrons in the narrow-gap Mott insulator VO$_2$. The initial stages of relaxation are treated using a quantum Boltzmann equation methodology, which reveals a rapid ($\sim$ femtosecond time scale) relaxation to a pseudothermal state characterized by a few parameters that vary slowly in time. The long-time limit is then studied by a Hartree-Fock methodology, which reveals the possibility of nonequilibrium excitation to a new metastable $M_1$ metal phase that is qualitatively consistent with a recent experiment. The general physical picture of photoexcitation driving a correlated electron system to a new state that is not accessible in equilibrium may be applicable in similar materials.
\end{abstract}

\maketitle

\section{Introduction}

Materials with strongly correlated electrons are of continuing interest and importance to condensed matter physics because of the remarkable variety of electronic phases they exhibit~\cite{Imada98,Kotliar04,Dagotto05,Takagi10}. It is now possible to experimentally drive these materials into strongly nonequilibrium states by, for example, the application of high intensity laser pulses~\cite{Norris03,Basov11}. These experiments raise questions of the dynamics of the photoexcited electrons and, in particular, they open the possibility of driving the system into a new phase \cite{Ao06} that may not be attainable in equilibrium. In this paper, we address these issues theoretically by developing a formalism to determine the evolution of the electron distribution function after the initial laser pulse and use a nonequilibrium Hartree-Fock formalism to investigate the possibility of new phases. The specific context of our work is a recent experiment on vanadium dioxide (VO$_2$) \cite{Morrison14}, which found that at certain levels of laser excitation, the low-temperature insulating phase could be driven into a long-lived metallic phase with no equilibrium analogue. We expect our theoretical methods and the general features of our results to be applicable to other systems as well.

Vanadium dioxide (VO$_2$) is a strongly correlated material. At temperatures $>340$~K, it exists in a metallic rutile ($R$) phase. At $\sim\,$340~K, VO$_2$ undergoes a first-order transition \cite{Morin59} to a monoclinic insulating phase with an optical band gap of $\sim 0.6$~eV \cite{Ladd69}. Two monoclinic phases have been reported \cite{Villeneuve73,Pouget76}. The one of interest here is the $M_1$ phase characterized by dimerized chains of V ions \cite{Andersson56}. Density functional theory plus dynamical mean field theory (DFT+DMFT) \cite{Kotliar06} calculations indicate that the insulating behavior of the $M_1$ phase arises from the combination of structural (dimerization) and electronic correlation effects, in the sense that calculations performed either without interaction or without the structural distortion predict metallic behavior \cite{Biermann05,Belozerov12}.

Very recently, Morrison \textit{et al.}~\cite{Morrison14}~performed pump-probe experiments, which uncovered new phenomena. In the experiments, a laser pulse was used to pump a density of electrons above the insulating gap. After the excitation, the time dependences of the electron diffraction spectrum and the mid-infrared (mid-IR) transmissivity were measured. For intermediate laser fluence, the mid-IR transmissivity decayed to zero on a sub-picosecond time scale, and then remained zero for the duration of the measurement ($>100$~ps), indicating the formation of a long-lived metallic state. However, the electron diffraction pattern showed that the system remained in the $M_1$ (dimerized) structure, the lattice structure normally associated with the insulating phase. Understanding this $M_1$ metal phase is a key goal of our work.

In this paper, we argue that photoexcitation can lead to the formation of a long-lived electronic state that is qualitatively different from the equilibrium phases found in the material. We first present an analysis of the time evolution of the electron distribution function soon after the incidence of an intense laser pulse, using realistic orbital and interaction structures of VO$_2$ in the $M_1$ phase. We assume that the decay of electron energy into phonons may be neglected and that the lattice does not have time to relax. We find that electrons rapidly ($\sim$~fs) relax to a pseudothermal distribution characterized by a single temperature but different chemical potentials of electrons and holes, which are determined by the input energy initially absorbed from the laser pulse and the number of electron-hole pairs initially photoexcited. The pseudothermal distribution then evolves over a time scale of $\sim\!10^2$~fs to a thermal distribution characterized by a unique temperature and chemical potential, which are determined by the input energy only. 

We then study the fate of the photoexcited system over longer times ($\sim$~ps) as the electron distribution cools. We show that for reasonable parameters, the thermal Hartree-Fock energy landscape exhibits two minima, one equilibrium state and one metastable state with different V-$3d$ orbitals preferentially occupied. This occupancy-dependent shift of the electron bands is a key ingredient in the electronic structure of strongly correlated materials. The metastable state is metallic for reasonable values of the interaction parameters. It is reachable within an input energy comparable to the pump laser fluence and not too much greater than the estimated absorption in the experimental situation in Ref.~\cite{Morrison14}.

The rest of the paper is organized as follows. In Sec. \ref{sec:DFT+U+V}, we present the band model that we will use. In Sec.~\ref{sec:Formalism:fixed-band}, we develop a quantum Boltzmann picture of the evolution of the nonequilibrium electron distribution. In Sec.~\ref{sec:Formalism:soft-band}, we provide a Hartree-Fock analysis of the circumstances under which a change in orbital occupancy can drive a change in electronic state. In Secs.~\ref{sec:Results}--\ref{sec:Results:soft-band}, we show the results obtained by applying our theory to VO$_2$ in Morrison's experiment \cite{Morrison14}. Section \ref{sec:Conclusion} is a summary and conclusion.

\section{The density functional+$U$+$V$ method for VO$_2$ \label{sec:DFT+U+V}}
Following Campo \textit{et al} \cite{Campo10}, we construct an electronic band structure for VO$_2$ using the density functional theory (DFT)+$U$+$V$ method, in which the basic density functional theory is supplemented by a Hartree-Fock treatment of the on-site (``$+U$'') and inter-site (``$+V$'') $d$--$d$ interactions. Belozerov \textit{et al} have constructed a DFT+DMFT+$V$ theory with very similar physics \cite{Belozerov12}. The effects of the ``$+V$'' term are a reasonable representation of the inter-site self-energy terms found in the cluster DMFT calculations of Biermann \textit{et al} \cite{Biermann05}. Note that in the correct orbital basis, these inter-site self-energy terms have only a weak frequency dependence \cite{Tomczak07}. Let us write the Kohn-Sham Hamiltonian of the electrons in their ground state as \cite{Campo10,Liechtenstein95}
\begin{align}
\hat{H}_0=\hat{H}_\mathrm{DFT}+\hat{\mathcal{V}}_\mathrm{HF}-\hat{H}_\mathrm{dc},
\label{eq:H0}
\end{align}
where $\hat{H}_\mathrm{DFT}$ comes from a density functional band calculation, $\hat{\mathcal{V}}_\mathrm{HF}$ is the Hartree-Fock approximation to the electron-electron interactions $\hat{\mathcal{V}}$ involving the vanadium $3d$ orbitals, and $\hat{H}_\mathrm{dc}$ is the double-counting correction.

In the $M_1$ phase of VO$_2$, the unit cell contains four vanadium ions, which form two dimerized pairs. We only consider interactions within one unit cell. These may be generally written as
\begin{align}
\hat{\mathcal{V}}=\frac{1}{2}\sum_{\vec{R}\sigma\sigma'}\sum_{\{m\}}U_{m_1\ldots m_4}\hat{c}_{\vec{R}m_1\sigma}^\dagger\hat{c}_{\vec{R}m_2\sigma'}^\dagger\hat{c}_{\vec{R}m_4\sigma'}\hat{c}_{\vec{R}m_3\sigma},
\label{eq:Ham-int}
\end{align}
where $\vec{R}$ labels the unit cells, $m_1\ldots m_4$ run over the correlated orbitals in a unit cell, and $\sigma,\sigma'$ label the spins. We consider two contributions to $\hat{\mathcal{V}}$: the on-site intra-$3d$ interactions, which we take to be the rotationally invariant form \cite{Liechtenstein95} including both $t_{2g}$ and $e_g$ orbitals parametrized by the Hubbard $U$ and Hund's coupling $J$, and inter-site interactions between the two vanadium ions in each dimer. The Hartree-Fock approximation $\hat{\mathcal{V}}_\mathrm{HF}$ of the electron-electron interactions $\hat{\mathcal{V}}$ takes the form
\begin{align}
\hat{\mathcal{V}}_\mathrm{HF}=\sum_{\vec{R}}\sum_{m_1m_2\sigma}V_{m_1m_2}\hat{c}^\dagger_{\vec{R}m_1\sigma}\hat{c}_{\vec{R}m_2\sigma},
\label{eq:V-HF}
\end{align}
where in a non-spin-polarized system (like VO$_2$)
\begin{align}
V_{m_1m_2}&=\sum_{m_3m_4\sigma'}(U_{m_1m_3m_2m_4}-U_{m_1m_3m_4m_2}\delta_{\sigma\sigma'})n_{m_4m_3}
\nonumber\\
&=\sum_{m_3m_4}(2U_{m_1m_3m_2m_4}-U_{m_1m_3m_4m_2})n_{m_4m_3}
\label{eq:V-matrix}
\end{align}
and the occupation matrix
\begin{align}
n_{m_4m_3}=\langle\hat{c}^\dagger_{\vec{R}m_3\sigma'}\hat{c}_{\vec{R}m_4\sigma'}\rangle
\label{eq:Occ-Matrix}
\end{align}
are independent of both spin and unit cell coordinate $\vec{R}$.

In Eq.~\eqref{eq:V-matrix}, $V_{m_1m_2}$ has both the on-site and inter-site intra-dimer terms. The on-site terms are the usual ones treated in standard DFT+$U$ calculations \cite{Liechtenstein95}. The inter-site terms are parametrized by a single parameter $V$ (the $V$ in DFT+$U$+$V$) and their contributions in $\hat{\mathcal{V}}_\mathrm{HF}$ take the form
\begin{align}
\hat{H}_V=-V\sum_{\vec{R}\sigma}\sum_{\langle m_1,m_2\rangle}\!n_{m_1m_2}\,\hat{c}_{\vec{R}m_1\sigma}^\dagger\hat{c}_{\vec{R}m_2\sigma},
\label{eq:HV}
\end{align}
\begin{figure}
\centering
\includegraphics[width=0.85\columnwidth]{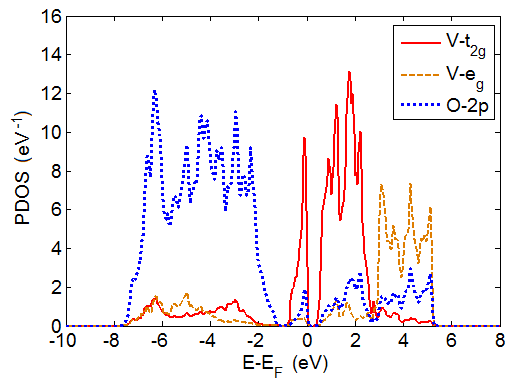}
\\(a)\\
\includegraphics[width=0.9\columnwidth]{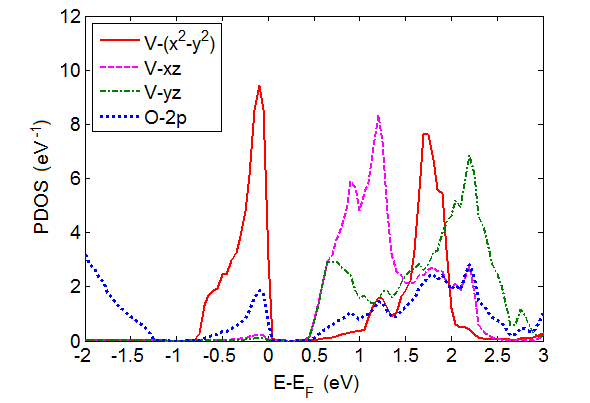}
\\(b)\\
\caption{(Color online) The projected density of states (PDOS) of the $M_1$ phase of VO$_2$ onto the maximally localized Wannier orbitals in DFT+$U$+$V$ in (a) the whole $p-d$ subspace and (b) the near-Fermi-level regime, with $U=4$~eV, $J=0.65$~eV, and $V=1$~eV. The three $d$-orbitals in (b) are defined as the $t_{2g}$-subspace because of the detailed crystal structure of VO$_2$. (See e.g.~Fig.~5 in \cite{Eyert02})
\label{fig:M1-insulator}}
\end{figure}

\noindent
which contains only the Fock terms of the density-density interaction $V\hat{n}_{\vec{R}m_1\sigma}\hat{n}_{\vec{R}m_2\sigma}$. The inter-site Hartree terms are assumed to be already included in $\hat{H}_\mathrm{DFT}$ and are not included again in $\hat{H}_V$ \cite{Campo10}. The Fock terms are orbitally diagonal, meaning that the $m_1$ and $m_2$ sum over only $d$ orbitals of the same type (e.g.~$d_{x^2-y^2}$--$d_{x^2-y^2}$, $d_{xz}$--$d_{xz}$, etc.)~in the two vanadium ions in a dimer. The inter-site matrix element $n_{m_1m_2}$ (hybridization) between different types of $d$ orbitals is typically small. In the ground state (insulating $M_1$ phase), only the hybridization of $d_{x^2-y^2}$ orbitals makes an appreciable contribution to $\hat{H}_V$, but in nonequilibrium metastable states, hybridizations of other $d$ orbitals may be also important, so we will keep the terms of all five $d$ orbitals in $\hat{H}_V$.

To proceed in practice, we first performed a non-spin-polarized DFT+$U$ calculation \cite{Liechtenstein95} using the Vienna Ab initio Simulation Package (VASP) \cite{Kresse96} with the atomic positions fixed in the experimental $M_1$ structure \cite{Andersson56}. We used a k-point mesh of $10\times 10\times 10$, an energy cutoff of 600~eV, and the projector-augmented wave Perdew-Burke-Ernzerhof (PAW-PBE) pseudopotential \cite{Perdew96} in the VASP library. The on-site interactions are parametrized by $U=4$~eV and $J=0.65$~eV \cite{Miyake09}. The $\hat{H}_\mathrm{DFT}$ in Eq.~\eqref{eq:H0} is then defined as the projection of the DFT+$U$ Hamiltonian onto a basis obtained from a Wannier fit to the 24 O-$2p$ and 20 V-$3d$ orbitals using Wannier90 \cite{Mostofi08} but with the on-site contributions to $V_{m_1m_2}$ and the double-counting terms removed. These on-site contributions plus the inter-site Fock terms $\hat{H}_V$ in Eq.~\eqref{eq:HV} then make up the remaining terms in Eq.~\eqref{eq:H0}.

The DFT+$U$+$V$ band structure for VO$_2$ is plotted in Fig.~\ref{fig:M1-insulator} for $V=1$~eV. The results are in good agreement with pre-existing results obtained using the GW method \cite{Continenza99} and cluster dynamical mean-field theory (CDMFT) \cite{Biermann05,Tomczak07}. The validity of modeling VO$_2$ in a renormalized band picture is corroborated in \cite{Tomczak07}. We will discuss our results in detail in Sec.~\ref{sec:Results:prep}.

\section{Formalism}
\subsection{Electron dynamics: initial relaxation \label{sec:Formalism:fixed-band}}

In this subsection, we present the formalism we use to study the initial relaxation of the photoexcited electrons before energy dissipates into other slower degrees of freedom such as phonons. For simplicity we use the quantum Boltzmann equation (QBE) \cite{Snoke10}, a dynamical equation for the occupancies $n_{\vec{k}\nu\sigma}$ of the Bloch states $|\vec{k}\nu\sigma\rangle$ in an electronic band structure, e.g.,
\begin{align}
\hat{H}_0=\sum_{\vec{k}\nu\sigma}\epsilon_{\vec{k}\nu}\hat{c}_{\vec{k}\nu\sigma}^\dagger\hat{c}_{\vec{k}\nu\sigma},
\end{align}
where $\hat{H}_0$ is the DFT+$U$+$V$ Hamiltonian in Eq.~\eqref{eq:H0}, $\vec{k}$ sums over k-points in the first Brillouin zone, $\nu$ is the band index and $\sigma$ labels the spin. The Kohn-Sham eigenvalues $\epsilon_{\vec{k}\nu}$ do not carry a spin index $\sigma$ in a non-spin-polarized system like VO$_2$.

The quantum Boltzmann equation treats electron-electron interactions $\hat{\mathcal{V}}$ via Fermi's golden rule, which gives the transition rates due to $\hat{\mathcal{V}}$ between different Slater-determinant eigenstates of $\hat{H}_0$. While this perturbative method fails to capture important aspects of correlated electrons, we believe that the main conclusions of this section, namely the order of magnitude of the time scales and the qualitative features of the resulting orbital distributions should be reasonable.

In a non-spin-polarized system, the quantum Boltzmann equation (QBE) is given by
\begin{align}
\frac{dn_{\vec{k}_1\nu_1}}{dt}=&\,\frac{2\pi}{\hbar}\frac{1}{N^2}\sum_{\vec{k}_2\vec{k}_3\vec{k}_4}\sum_{\nu_2\nu_3\nu_4}|\tilde{U}_{\nu_1\nu_2\nu_3\nu_4}(\vec{k}_1\vec{k}_2\vec{k}_3\vec{k}_4)|^2
\nonumber\\
&\times\delta_{\vec{k}_1+\vec{k}_2,\vec{k}_3+\vec{k}_4}\delta(\epsilon_{\vec{k}_1\nu_1}\!\!+\epsilon_{\vec{k}_2\nu_2}\!\!-\epsilon_{\vec{k}_3\nu_3}\!\!-\epsilon_{\vec{k}_4\nu_4})
\nonumber\\
&\times\left[(1-n_{\vec{k}_1\nu_1})(1-n_{\vec{k}_2\nu_2})n_{\vec{k}_3\nu_3}n_{\vec{k}_4\nu_4}\right.
\nonumber\\
&\left.\quad-n_{\vec{k}_1\nu_1}n_{\vec{k}_2\nu_2}(1-n_{\vec{k}_3\nu_3})(1-n_{\vec{k}_4\nu_4})\right],
\label{eq:QBE-std}
\end{align}
where $N$ is the total number of k-points, and the matrix element $|\tilde{U}_{\nu_1\nu_2\nu_3\nu_4}(\vec{k}_1\vec{k}_2\vec{k}_3\vec{k}_4)|^2$ is a short-hand symbol for $|\langle\vec{k}_1\nu_1\sigma,\vec{k}_2\nu_2\sigma'|\hat{\mathcal{V}}|\vec{k}_3\nu_3\sigma,\vec{k}_4\nu_4\sigma'\rangle|^2$ summed over the $\sigma=\sigma'$ and $\sigma\neq\sigma'$ cases. The occupancies $n_{\vec{k}\nu}=n_{\vec{k}\nu\uparrow}=n_{\vec{k}\nu\downarrow}$ are single-spin quantities. The k-variables sum over only the first Brillouin zone and the Kronecker $\delta_{\vec{k}_1+\vec{k}_2,\vec{k}_3+\vec{k}_4}$ is to be interpreted as implying equivalence up to a reciprocal lattice vector to correctly impose the conservation of crystal momentum.

A direct simulation of Eq.~\eqref{eq:QBE-std} in a general band structure is numerically difficult. The main problem comes from the energy delta function, which requires $\epsilon_{\vec{k}_1\nu_1}\!\!+\,\epsilon_{\vec{k}_2\nu_2}\!=\epsilon_{\vec{k}_3\nu_3}\!\!+\,\epsilon_{\vec{k}_4\nu_4}$. To ensure the conservation of energy to the needed accuracy, one has to choose a very dense k-point mesh, which then leads to too many degrees of freedom to handle. In order to obtain a computationally tractable model that still captures the important physics, we construct a momentum-averaged quantum Boltzmann equation, whose key variables are the energy distributions of electrons in different bands without any k-point information. Let us begin the derivation by averaging the matrix elements $|\tilde{U}_{\nu_1\nu_2\nu_3\nu_4}(\vec{k}_1\vec{k}_2\vec{k}_3\vec{k}_4)|^2$ over the 4 k-variables to define
\begin{align}
\overline{|U|^2}_{\nu_1\nu_2\nu_3\nu_4}=&\sum_{\{\vec{k}\}}|\tilde{U}_{\nu_1\nu_2\nu_3\nu_4}(\vec{k}_1\vec{k}_2\vec{k}_3\vec{k}_4)|^2
\delta_{\vec{k}_1+\vec{k}_2,\vec{k}_3+\vec{k}_4}\nonumber\\
&\times\delta(\epsilon_{\vec{k}_1\nu_1}\!\!+\epsilon_{\vec{k}_2\nu_2}\!\!-\epsilon_{\vec{k}_3\nu_3}\!\!-\epsilon_{\vec{k}_4\nu_4})
\nonumber\\
&\Big/\frac{1}{N}
\sum_{\{\vec{k}\}}\delta(\epsilon_{\vec{k}_1\nu_1}\!\!+\epsilon_{\vec{k}_2\nu_2}\!\!-\epsilon_{\vec{k}_3\nu_3}\!\!-\epsilon_{\vec{k}_4\nu_4}),
\label{eq:U2-kavg}
\end{align}
which are the k-averaged matrix elements that only depend on the 4 band indices $\nu_1\ldots\nu_4$. The motivation for the k-averaging comes from the local nature of the interaction $\hat{\mathcal{V}}$ defined in Eq.~\eqref{eq:Ham-int}. The k-dependence of $|\tilde{U}_{\nu_1\nu_2\nu_3\nu_4}(\vec{k}_1\vec{k}_2\vec{k}_3\vec{k}_4)|^2$ comes purely from the Bloch wave functions and tends to be complicated, and effectively random in real materials, so averaging over the momentum variables is reasonable. Next, we assume that the occupation numbers of the Bloch states
\begin{align}
n_{\vec{k}\nu}\approx n_{\nu}(\epsilon_{\vec{k}\nu})
\label{eq:occ-num}
\end{align}
are only functions of band index $\nu$ and energy $\epsilon_{\vec{k}\nu}$. Then defining the single-spin density of states of band $\nu$
\begin{subequations}
\begin{align}
D_\nu(E)=\frac{1}{N}\sum_{\vec{k}}\delta(\epsilon_{\vec{k}\nu}-E),
\label{eq:density-of-states}
\end{align}
and the densities of occupied and empty states
\begin{align}
N_\nu(E)&=D_\nu(E)n_\nu(E),
\label{eq:density-of-electrons}\\
\bar{N}_\nu(E)&=D_\nu(E)[1-n_\nu(E)],
\label{eq:density-of-holes}
\end{align}
\end{subequations}
we derive a k-averaged QBE
\begin{align}
\frac{dN_{\nu_1}(E_1)}{dt}&=\frac{2\pi}{\hbar}\sum_{\nu_2\nu_3\nu_4}\overline{|U|^2}_{\nu_1\nu_2\nu_3\nu_4}\int dE_2dE_3dE_4
\nonumber\\
&\times\delta(E_1+E_2-E_3-E_4)
\nonumber\\
&\times\left[\bar{N}_{\nu_1}(E_1)\bar{N}_{\nu_2}(E_2)N_{\nu_3}(E_3)N_{\nu_4}(E_4)\right.
\nonumber\\
&\left.\;\;-N_{\nu_1}(E_1)N_{\nu_2}(E_2)\bar{N}_{\nu_3}(E_3)\bar{N}_{\nu_4}(E_4)\right].
\label{eq:QBE-kavg}
\end{align}
The band indices are kept in full. The \textit{ab initio} rate constants $\overline{|U|^2}_{\nu_1\nu_2\nu_3\nu_4}$ are obtained from Eq.~\eqref{eq:U2-kavg} using Monte Carlo methods on a Wannier interpolated k-point mesh of $20\times 20\times 20$. We will give a detailed derivation of Eq.~\eqref{eq:QBE-kavg} in Appendix \ref{sec:appendix-A}.

\subsection{Soft bands in Hartree-Fock theory \label{sec:Formalism:soft-band}}
In density function theory, the electronic potential is a self-consistently determined functional of the electron density, so that changes in the electron distribution will lead to changes in the band structure. This effect is greatly enhanced in extended DFT theories such as DFT+$U$ and DFT+$U$+$V$ because, in particular, the relative energetics of the different $d$ orbitals depends strongly on the orbital occupation matrix. This strong dependence may lead to photoinduced phase transitions if photoexcitation changes the occupancy sufficiently.

In the specific case of VO$_2$, since the wavelength of the pump laser is typically $800$~nm ($E_\mathrm{photon}=1.55$~eV), the pump laser typically changes the electron distribution among the V-$3d$ orbitals (see Fig.~\ref{fig:M1-insulator}), but does not change the total $d$-count or the real-space charge density $n(\mathbf{r})$ significantly. We therefore argue that we may analyze the effects of photoexcitation using Eq.~\eqref{eq:H0} with $\hat{H}_\mathrm{DFT}$ and $\hat{H}_\mathrm{dc}$ left unchanged, but with $\hat{\mathcal{V}}_\mathrm{HF}$ now determined by the nonequilibrium distribution of electrons over orbitals, i.e., the Kohn-Sham Hamiltonian becomes
\begin{align}
\hat{H}=\hat{H}_0+\Delta\hat{\mathcal{V}}_\mathrm{HF},
\label{eq:H-soft-band}
\end{align}
where $\Delta\hat{\mathcal{V}}_\mathrm{HF}$ is the change of $\hat{\mathcal{V}}_\mathrm{HF}$ due to the change of the orbital occupation matrix (see Eqs.~\eqref{eq:V-HF}--\eqref{eq:Occ-Matrix}) under photoexcitation.

Eq.~\eqref{eq:H-soft-band} implies that the electronic band structure becomes soft in the sense that the conduction band floats down when its occupancy increases and the valence band floats up when its occupancy decreases under photoexcitation. This general picture shows that photo-excitation has the potential of closing the Mott gap and driving an insulator-metal transition, thus giving rise to new electronic phases. The total energies of different electronic states can be compared using
\begin{align}
E_\mathrm{tot}=\langle\hat{H}\rangle-\frac{1}{2}\langle\hat{\mathcal{V}}_\mathrm{HF}\rangle+\mathrm{const},
\label{eq:Etot}
\end{align}
where the expectation value is now taken using the nonequilibrium distribution. In Sec.~\ref{sec:Results:soft-band}, we will use Eq. \eqref{eq:Etot} to construct an energy landscape for nonequilibrium VO$_2$ that will be used to interpret the experiments of Morrison \textit{et al} \cite{Morrison14}.

\section{Results \label{sec:Results}}
\subsection{Band structure and initial excitation \label{sec:Results:prep}}

Using the DFT+$U$+$V$ method outlined in Sec.~\ref{sec:DFT+U+V} with $U=4$~eV, $J=0.65$~eV and $V=1$~eV, we obtain the band structure of the $M_1$ phase of VO$_2$ in its ground state shown in Fig.~\ref{fig:M1-insulator}. For these parameter values, the optical gap at the Fermi level is 0.62~eV (not plotted) in good agreement with experiment \cite{Ladd69}, and the indirect gap between the highest occupied and lowest unoccupied Bloch states (which we will later call the HOMO-LUMO gap) is 0.45~eV. The lower gap separating the V-$3d$ and O-$2p$ dominant bands below the Fermi level is 0.55~eV. We note, however, that the range of parameters $U=3.5$ $\sim 4.5$~eV and correspondingly $V=1.4\sim 0.6$~eV provide equally reasonable descriptions of the material.

The band structure in Fig.~\ref{fig:M1-insulator} is characterized by substantial orbital ordering, with the bonding $d_{x^2-y^2}$ orbital highly occupied and all other $d$ orbitals nearly empty. The bonding-antibonding splitting of the $d_{x^2-y^2}$ orbitals arises from the dimerization of the crystal structure and is enhanced by the inter-site Fock interaction $V$. The bonding-antibonding transition is optically active and the energy $\sim 2$~eV in Fig.~\ref{fig:M1-insulator} is in agreement with optical conductivity data \cite{Qazilbash08}. The HOMO-LUMO gap lies in between the $d_{x^2-y^2}$ bonding orbital and the $d_{xz}$ and $d_{yz}$ orbitals (which do not exhibit a significant bonding-antibonding splitting). The gap is opened due in large part to the on-site Coulomb interaction $U$. If $\,U=0$, the $d_{xz}$ and $d_{yz}$ orbitals would overlap the $d_{x^2-y^2}$ orbital in energy and a metallic state would result. If $V=0$, $U=4$~eV, the antibonding $d_{x^2-y^2}$ peak would fall to the bottom of the conduction band and the HOMO-LUMO gap would decrease to only 0.13~eV (not shown).

Next we estimate the energy range and number of electrons photoexcited in a recent pump-probe experiment on VO$_2$ \cite{Morrison14}. The wavelength of the pump laser is $\lambda=800$~nm ($E_\mathrm{photon}=1.55$~eV). Inspection of Fig.~\ref{fig:M1-insulator} indicates that the primary excitation is from the $d_{x^2-y^2}$ bonding band to $d_{xz}$-derived states. Solving the optics problem for the experimental geometry specified in \cite{Morrison14} reveals that the experimental fluence of $3.7\sim 9$~mJ/cm$^2$ that yielded an $M_1$ metal initially generates $N_\mathrm{eh}^0=0.048\sim 0.12$ electron-hole pairs per unit cell, corresponding to an energy increase per unit cell of $\Delta E_\mathrm{tot}=0.074\sim 0.18$~eV (see Appendix \ref{sec:appendix-B} for details).

\subsection{Fixed-band QBE dynamics \label{sec:Results:fixed-band}}

\begin{figure}[b]
\includegraphics[width=\columnwidth]{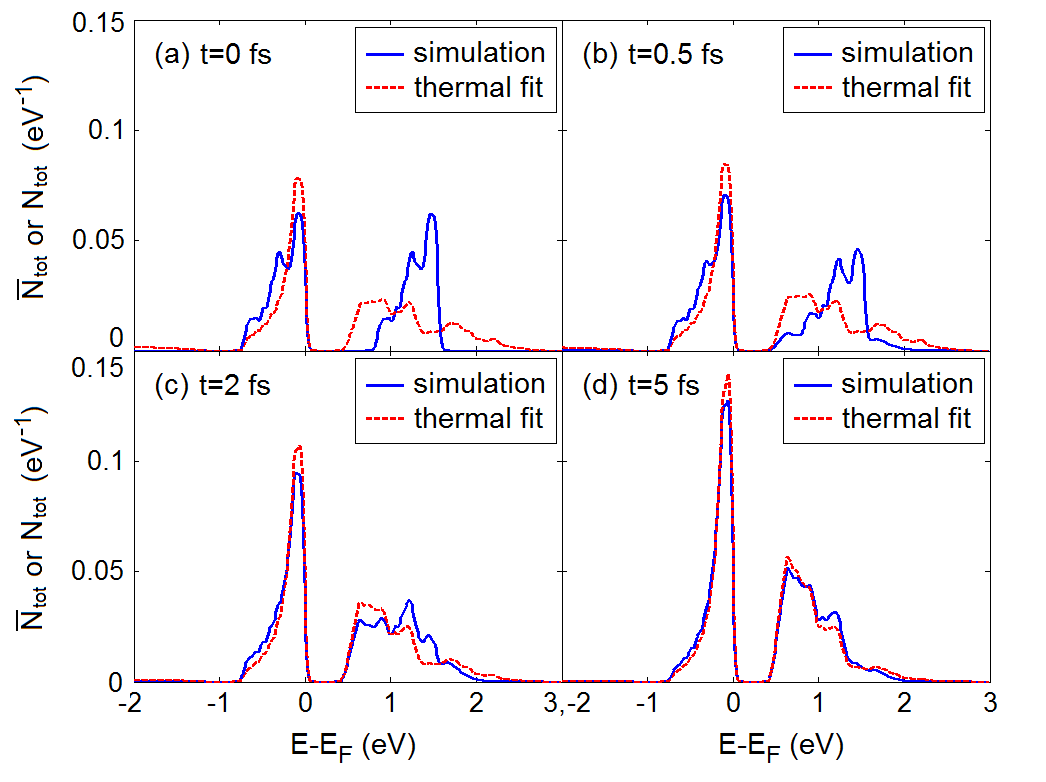}
\caption{(Color online) The hole distribution $\bar{N}_{\mathrm{tot}}(E)$ ($E\!<\!E_F$) and electron distribution $N_{\mathrm{tot}}(E)$ ($E\!>\!E_F$) per spin at (a) $t\!=\!0$~fs, (b) $t\!=\!0.5$~fs, (c) $t\!=\!2$~fs, and (d) $t\!=\!5$~fs. Laser fluence$\,=3.7$~mJ/cm$^2$. The distribution is fitted to a Fermi distribution with a common temperature $T$ but two chemical potentials $\mu_e$ and $\mu_h$ for the electrons and holes based on the energy and the number of electron-hole pairs at every instant.\!\label{fig:NE-t}}
\end{figure}

\begin{figure}
\includegraphics[width=\columnwidth]{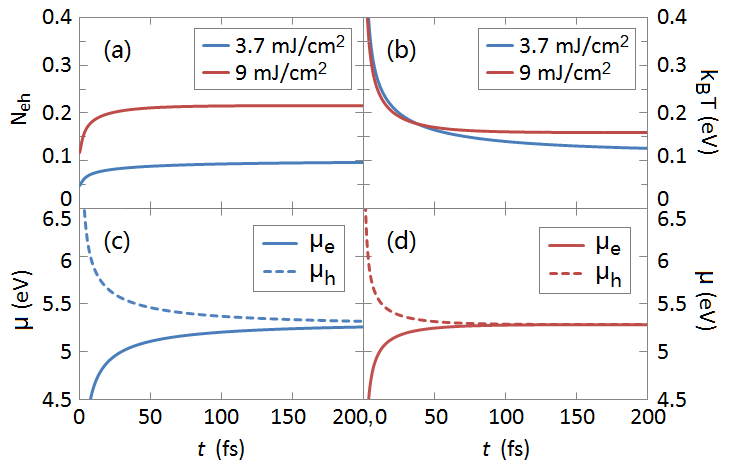}
\caption{(Color online) Time evolution of (a) number of electron-hole pairs $N_\mathrm{eh}$ per unit cell, (b) temperature $T$, (c) chemical potentials $\mu_e$ and $\mu_h$ under laser fluence $=3.7$~mJ/cm$^2$ and (d) $9$~mJ/cm$^2$. \label{fig:thermal-params}}
\end{figure}

In this subsection, we simulate the evolution of the electron distribution immediately after the initial laser pulse using the quantum Boltzmann equation methods of Sec.~\ref{sec:Formalism:fixed-band} over a fixed band structure in Fig.~\ref{fig:M1-insulator}. We begin by assuming for simplicity that the absorption is proportional to the product of densities of states at energy separation $\hbar\omega=1.55$~eV. Then at $t=0$, immediately after the laser pulse, we have the distributions of holes and electrons given by
\begin{align}
\bar{N}_{\mathrm{tot}}(E)=N_{\mathrm{tot}}(E+\hbar\omega)\propto D_\mathrm{tot}(E)D_\mathrm{tot}(E+\hbar\omega),
\!\!\!\!\!\phantom{\frac{1}{2}}
\label{eq:init-distr}
\end{align}
where $E$ satisfies $E<E_F$ and $E+\hbar\omega-E_F>0.45$~eV, the HOMO-LUMO gap. Here the subscript ``tot'' means to sum over all bands $\nu$. The total number of electron-hole pairs $N_\mathrm{eh}^0$ is determined by the experimental laser fluence, as discussed at the end of Sec.~\ref{sec:Results:prep}. Then we assume that the initially excited electrons and holes are randomly distributed over band states, i.e., for all energy $E$, the density of occupied states in band $\nu$,
\begin{align}
N_\nu(E)=\frac{D_\nu(E)}{D_\mathrm{tot}(E)}N_\mathrm{tot}(E),\!\!\!\phantom{\begin{matrix}
\ \\ \ \\ \
\end{matrix}}
\end{align}
is directly proportional to the density of states $D_{\nu}(E)$ in band $\nu$. We then evolve the distribution according to Eq.~\eqref{eq:QBE-kavg}. We find that the equilibration process comes in basically two steps: the fast prethermalization (Fig.~\ref{fig:NE-t}) that establishes a pseudothermal distribution characterized by a common temperature $T$ but different chemical potentials $\mu_e$ and $\mu_h$ for the electrons and holes, and then the slow evolution of thermal parameters $T,\mu_e,\mu_h$ (Fig.~\ref{fig:thermal-params}) to the final thermal state.

Fig.~\ref{fig:NE-t} shows the initial stages of relaxation for laser fluence $=3.7$~mJ/cm$^2$, comparing the calculated distribution to the distribution expected if the electrons and holes have thermalized. In the first $\sim0.5$~fs after the laser pulse, the distribution of photoexcited electrons develops a tail to both high and low energies. Then in the next $1\sim 2$ femtoseconds, the electron and hole distributions thermalize. At the same time, the number of electrons and holes begins to increase due to the inverse Auger process, in which a high-energy electron scatters to a low-energy state by creating an electron-hole pair, thereby increasing the electron and hole densities and shifting the main weight in the conduction band to lower energies (a similar effect was noted in the Hubbard model by Eckstein and Werner \cite{Eckstein11}). However, as the electrons thermalize, the inverse Auger scattering rate decreases rapidly since only electrons far out in the tail of the pseudothermal distribution have enough energy to down-scatter to create an electron-hole pair while still remaining in the conduction band. By $t=5$~fs, the electron and hole distributions are fully thermalized and the subsequent evolution can be described by the evolution of thermal parameters. For higher laser fluence $=9$~mJ/cm$^2$ (not shown) the time evolution is qualitatively the same as in Fig.~\ref{fig:NE-t} and takes roughly the same time, but produces more electron-hole pairs (Fig.~\ref{fig:thermal-params}).

The evolution of thermal parameters, i.e., the temperature $T$, the chemical potential $\mu_e$ of the electrons and $\mu_h$ of the holes, is much slower as noted above. Fig.~\ref{fig:thermal-params} shows the results for both the low fluence $3.7$~mJ/cm$^2$ and the high fluence $9$~mJ/cm$^2$. The equilibration time constant approximately scales as the inverse of the square of the number of electron-hole pairs $N_\mathrm{eh}$ at equilibrium, which is a signature of the three-particle Auger and inverse Auger scattering processes.

Much of what happens in the simulation are explained by the rate constants $\overline{|U|^2}_{\nu_1\nu_2\nu_3\nu_4}$. The largest rate constants are those of the hole-hole, electron-hole, and electron-electron scattering processes that do not change $N_\mathrm{eh}$. The pair creation and recombination processes that change $N_\mathrm{eh}$ are comparatively slow. This separation of time scales has two origins: (a) the gap, which means that the processes must involve electrons in the tail of the distribution, and (b) the different orbital characters of the top of the valence band ($d_{x^2-y^2}$) and the bottom of the conduction band ($d_{xz}$ and $d_{yz}$) in Fig.~\ref{fig:M1-insulator}, which means that changes in $N_\mathrm{eh}$ must come from orbital-changing interactions, i.e., the pair hopping and exchange terms $\sim J$, which are much smaller than the orbitally diagonally interactions $\sim U$.

Even though the density relaxation is much slower than prethermalization, due to the combination of small matrix element and kinetic bottleneck, our QBE-based simulation still finds that electrons in VO$_2$ will equilibrate in hundreds of femtoseconds. The higher the laser fluence, the faster the electrons equilibrate, as shown in Fig.~\ref{fig:thermal-params}. Based on the qualitative picture described in Sec.~\ref{sec:Formalism:soft-band} that photoexcitation generally narrows or closes the gap, reducing the bottleneck effect of electron relaxation, we expect that the beyond-fixed-band effects will lead to even faster relaxation, and to a larger final number of excited particle-hole pairs.

\section{Nonequilibrium phase transition to a metastable metallic state \label{sec:Results:soft-band}}

In Sec.~\ref{sec:Results:fixed-band}, we showed that electrons in VO$_2$ relax on a sub-picosecond time scale to a thermal state with a well-defined instantaneous temperature. In this section, we investigate whether the changes in orbital occupancies due to photoexcitation can lead to significant changes in the band structure, in particular the HOMO-LUMO gap. Because the system relaxes rapidly to a thermal state, we can avoid solving a dynamical Hartree-Fock equation and consider a Hartree-Fock theory in thermal states only.

We note at the outset that obtaining an insulating state in VO$_2$ requires two effects. First, the dimerization (enhanced by an inter-site correlation effect) splits the $d_{x^2-y^2}$ band into bonding and antibonding portions. Second, the on-site interaction produces a level splitting between $d_{x^2-y^2}$ and the $d_{xz}/d_{yz}$ orbitals. The dimerization gives the possibility of having a filled band, and the level splitting ensures that the $d_{x^2-y^2}$ band lies far enough below the other bands that it is indeed fully occupied. The equilibrium phase transition from the insulating to the metallic state involves a change in the crystal structure, removing the dimerization. An alternative possibility is that at fixed structure a population inversion of the $d_{x^2-y^2}$ and the $d_{xz}/d_{yz}$ bands, driven by photoexcitation, would lead to a reversal of the energy ordering, so that the non- (weakly) dimerized $d_{xz}/d_{yz}$ bands would lie lowest, creating an $M_1$ metal phase.

\begin{figure}
\centering
\includegraphics[width=0.95\columnwidth]{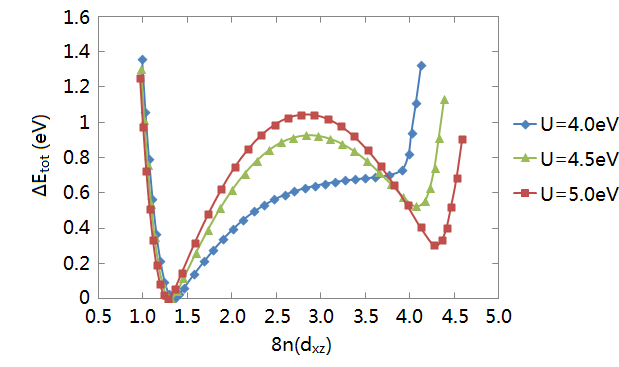}
\caption{(Color online) The energy landscape at different values of $U$ with $J=0.65$~eV and $V=1$~eV. The insulating phase is used as an energy reference point. The occupancy $n(d_{xz})$ is that per V ion per spin, and $8n(d_{xz})$ gives the number of V-$d_{xz}$ electrons per unit cell. $\Delta E_\mathrm{tot}$ is the total energy change per unit cell.
\label{fig:dEtot-ndxz}}
\end{figure}

To investigate the possibility of this $M_1$ metal phase, we first apply the Hartree-Fock theory in Sec.~\ref{sec:Formalism:soft-band} at temperature $T=0$ by calculating the shift of the bands using Eq.~\eqref{eq:H-soft-band}.~We start from an occupation matrix with a high $d_{xz}$ occupancy, and find at $\,U=4$~eV, $V=1$~eV and $J=0.65$~eV that our system relaxes back to the conventional $M_1$ insulator phase shown in Fig.~\ref{fig:M1-insulator} in the Hartree-Fock iterations. However, at slightly increased values of $\,U$, i.e., $\,U=4.5$~eV and $5$~eV, the iterations bring us to a new self-consistent state with a high $d_{xz}$ (low $d_{x^2-y^2}$) occupancy and no gap at the Fermi level -- an $M_1$ metal phase is found.

We next construct in Fig.~\ref{fig:dEtot-ndxz} a cut across the energy landscape as a function of orbital occupancies with the $M_1$ insulator and metal phases as its local minima. To do this, we first determine for $U=4.5$ and $5$~eV the $44\times 44$ (full $p$--$d$ basis) real-space density matrix of an intermediate state as a linear interpolation between the density matrices of the two local minima. Then we introduce k-independent Lagrange multipliers to the Kohn-Sham Hamiltonian $\hat{H}$, which are adjusted so that the band occupancies reproduce this interpolated density matrix. The states obtained are the minimum energy states subject to the constraint of a linearly interpolated real-space density matrix. The energy is then evaluated by Eq.~\eqref{eq:Etot} using $\hat{H}$ without the Lagrange multipliers. The resulting curve, although not necessarily the minimum energy path between the $M_1$ insulator and metal phases, should give a reasonable representation of the energy barrier between them. For $U=4$~eV, the metal phase is a state in the ghost region of the iterative Hartree-Fock dynamics with the slowest evolution, and the energy curve is plotted following the evolution to the insulating ground state. The extrapolated states at any value of $\,U$ cannot be obtained by linear extrapolation of real-space density matrices, as these can have occupancy eigenvalues not between 0 and 1. Instead, the states are obtained by tuning the orbital energies of $d_{xz}$ and $d_{yz}$ with respect to $d_{x^2-y^2}$ using the Lagrange multipliers to further raise or lower the $d_{xz}$ occupancy.

\begin{figure}
\centering
\includegraphics[width=0.85\columnwidth]{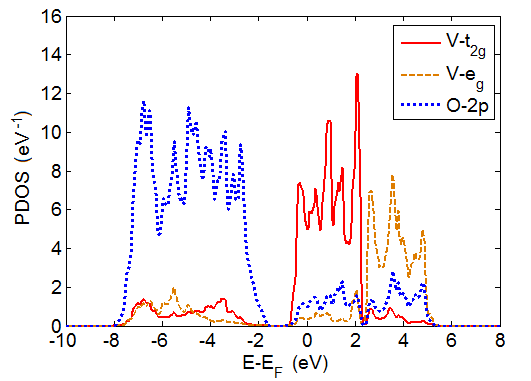}
\\(a)\\
\includegraphics[width=0.9\columnwidth]{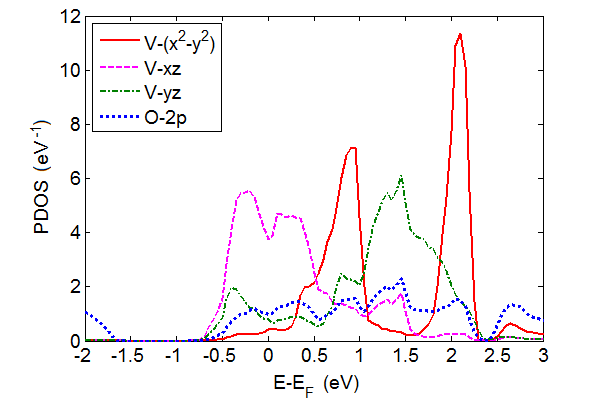}
\\(b)\\
\caption{(Color online) The projected density of states (PDOS) of the $M_1$ metal phase of VO$_2$ onto the maximally localized Wannier orbitals in DFT+$U$+$V$ in (a) the whole $p-d$ subspace and (b) the near-Fermi-level regime, with $U=4.5$~eV, $J=0.65$~eV, and $V=1$~eV. \label{fig:M1-metal}}
\end{figure}

The projected density of states of the $M_1$ metal phase is plotted in Fig.~\ref{fig:M1-metal}. We see that the density of states at the Fermi level is nonzero, so within a band picture the state is metallic. Also, the $d_{x^2-y^2}$ orbitals are now substantially above the Fermi level, and the bonding-antibonding splitting of the orbitals is less, reflecting the decrease in the inter-site Fock terms $\hat{H}_V$ due to the depletion of the $d_{x^2-y^2}$ band.

\begin{figure}
\centering
\includegraphics[width=0.9\columnwidth]{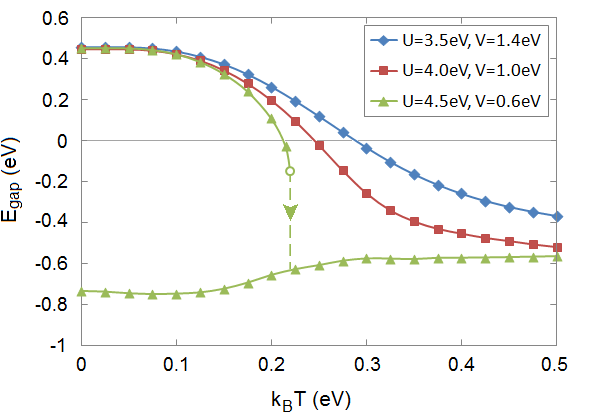}
\\(a)\\
\includegraphics[width=0.9\columnwidth]{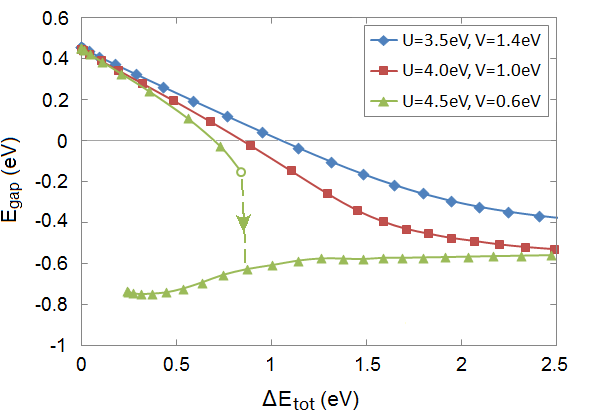}
\\(b)\\
\caption{(Color online) The energy gap $E_\mathrm{gap}$ v.s.~temperature $T$ and energy $\Delta E_\mathrm{tot}$ injected per unit cell by the laser pulse under various parameter values $U$ and $V$. The Hund's coupling $J=0.65$~eV is kept constant. Both the optical gap ($0.62$~eV, not plotted) and the HOMO-LUMO gap ($0.45$~eV) at $T=0$ are approximately kept constant under the simultaneous change of $U$ and $V$.\label{fig:Egap-T}}
\end{figure}

While Fig.~\ref{fig:dEtot-ndxz} shows that the $M_1$ metal phase has higher energy at $T=0$, we find that at $T>0$ the state may be favored. Fig.~\ref{fig:Egap-T} plots the calculated HOMO-LUMO gap as a function of the energy deposited by the pump laser into the sample for realistic parameter values discussed in Sec.~\ref{sec:Results:prep}. Because the electrons equilibrate rapidly, this is equivalent to plotting against temperature, although the temperature-energy relationship is not quite linear and depends on which phase the system is in.

Two qualitatively different behaviors are seen in Fig.~\ref{fig:Egap-T}. For $U=4$~eV, $V=1$~eV, there is no phase transition. The bonding $d_{x^2-y^2}$ band in Fig.~\ref{fig:M1-insulator} shifts up and the $d_{xz}$ and $d_{yz}$ bands shift down as temperature rises, and eventually the band gap between them is closed. But there is always a unique stable state at every temperature $T$ or energy $\Delta E_\mathrm{tot}$. Similar effects are seen for $U=3.5$~eV, $V=1.4$~eV except that the curve drops more slowly and the gap closes at a slightly higher temperature. The behavior is very different for $U=4.5$~eV, $V=0.6$~eV. When the overlap of the $d_{x^2-y^2}$ band with $d_{xz}$ and $d_{yz}$ bands (indicated by a negative gap in Fig.~\ref{fig:Egap-T}) exceeds a certain threshold (the small circle on the green curve), the band structure undergoes a first-order phase transition to a state with an inverted population and thus a negative HOMO-LUMO gap (metallic state) occurs. Near the discontinuity, the $E_\mathrm{gap}$--$T$ curve in Fig.~\ref{fig:Egap-T}a shows a $(T_c-T)^{1/2}$ singularity, but the $E_\mathrm{gap}$--$\Delta E_\mathrm{tot}$ curve in Fig.~\ref{fig:Egap-T}b is not singular.

\begin{figure}
\centering
\includegraphics[width=0.7\columnwidth]{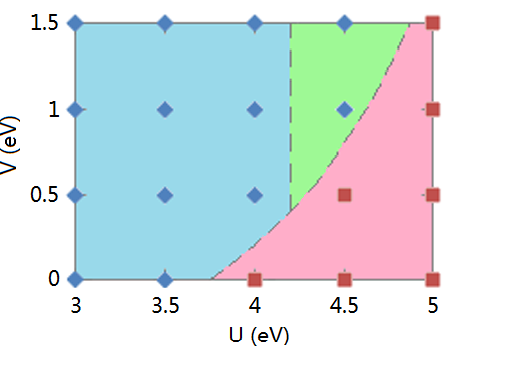}
\caption{(Color online) The $U$-$V$ phase diagram. At every blue diamond point, the $E_\mathrm{gap}$-$T$ curve is smooth, indicating a reversible insulator-metal transition. At every red square point, a discontinuity in $E_\mathrm{gap}$ occurs as temperature $T$ rises above a threshold and the system irreversibly jumps into a metal phase. The metal phase survives at $T=0$ the entire red and green regions but relaxes to the conventional insulating phase if parameters go back to the blue region.
\label{fig:UV-diagram}}
\end{figure}

The $M_1$ metal phase may be metastable (correspond to a local energy minimum) even if it is not thermally reachable. Fig.~\ref{fig:UV-diagram} summarizes the situation, showing by red squares (blue diamonds) the region where a thermally driven transition to the $M_1$ metal phase occurs (or not), and by colors (red and green) the regions where the $M_1$ metal phase is locally stable and (blue) where only the $M_1$ insulator phase is locally stable.

Compared with the input energy $\Delta E_\mathrm{tot}=0.074\sim 0.18$~eV per unit cell (4 VO$_2$) estimated in Sec.~\ref{sec:Results:prep} for the experiment in Ref.~\cite{Morrison14}, the transition point in Fig.~\ref{fig:Egap-T}b corresponds to a fluence about 4 times larger than that at which the putative $M_1$ metal phase was observed. At the experimental fluence level, the theory indicates that the HOMO-LUMO gap is only slightly reduced from $0.45$~eV in the insulating ground state to $0.35\sim 0.40$~eV (Fig.~\ref{fig:Egap-T}b, $U=4$~eV, $V=1$~eV). This discrepancy with experiment may be due to limitations of the Hartree-Fock theory, which does not calculate the energy of correlated electrons or locate phase boundaries accurately.

\section{Conclusion \label{sec:Conclusion}}
This paper presents a theoretical study of photoexcited VO$_2$ motivated by the recent experimental report \cite{Morrison14} of a long-lived metallic phase created by photoexcitation in a material with a crystal structure associated with insulating equilibrium behavior. We used a band-theory-based Hartree-Fock mean-field methodology combined with quantum Boltzmann equation treatment of excited-state kinetics. The key findings of our study were (a) very rapid ($\sim$~fs) relaxation of the photoexcited carriers to a pseudothermal state characterized by a common temperature but different chemical potentials for the electron and hole distributions, (b) a rapid ($\sim 10^2$~fs) relaxation to a thermal state with a well-defined common temperature and chemical potential and (c) the existence of a metallic phase which is metastable at temperature $T=0$ and can become favored at higher temperatures (or laser fluence levels).

The key approximations of our work are the Hartree-Fock plus Fermi's golden rule treatment of the electron-electron interactions, and neglect of electron-phonon coupling beyond thermal energy exchange. We believe that these approximations are not crucial. The important conclusion of the quantum Boltzmann and Fermi's golden rule studies of the dynamics is that thermalization of the excited particles proceeds much faster than experimental time scales, so that experimentally relevant issues, in particular the existence of a metastable metallic state, can be addressed using steady-state arguments. Further, the local stability of the metallic $M_1$ phase means that as phonons take energy out of the electronic system, the system may simply remain in this phase over a long time determined by nucleation kinetics. The conclusion seems very likely to survive the inclusion of higher order effects in the dynamics. Hartree-Fock theory is normally reliable for the identification of phases, although the estimates of the locations of phase boundaries may be inaccurate. The results presented here should be viewed as indicating the theoretical possibility of a metastable metallic phase for reasonable parameters. Further investigations of this metallic phase, including more reliable determination of the phase boundaries, investigation of the processes by which the metastable state might decay, and the study of the evolution of the lattice structure, would be of considerable interest.

\textit{Acknowledgments:} We thank P. Werner for helpful discussions. AJM acknowledges the warm hospitality and stimulating intellectual environment of the College de France, where part of the work on this manuscript was completed. This research is supported by the Department of Energy under grant DE-SC0012375.

\appendix
\section{The k-averaged QBE \label{sec:appendix-A}}
In this appendix, we give a detailed derivation of the k-averaged quantum Boltzmann equation (QBE) in Eq.~\eqref{eq:QBE-kavg} of the main text from the standard QBE in Eq.~\eqref{eq:QBE-std}. As one can see, when the assumption in Eq.~\eqref{eq:occ-num} is satisfied, the number of degrees of freedom of the system is greatly reduced and Eq.~\eqref{eq:QBE-std} becomes
\begin{align}
&\!\!
\frac{dn(\epsilon_{\vec{k}_1\nu_1})}{dt}=\frac{2\pi}{\hbar}\frac{1}{N^2}\sum_{\vec{k}_2\vec{k}_3\vec{k}_4}\sum_{\nu_2\nu_3\nu_4}|\tilde{U}_{\nu_1\nu_2\nu_3\nu_4}(\vec{k}_1\vec{k}_2\vec{k}_3\vec{k}_4)|^2
\nonumber\\
&\;\times\delta_{\vec{k}_1+\vec{k}_2,\vec{k}_3+\vec{k}_4}\delta(\epsilon_{\vec{k}_1\nu_1}\!\!+\epsilon_{\vec{k}_2\nu_2}\!\!-\epsilon_{\vec{k}_3\nu_3}\!\!-\epsilon_{\vec{k}_4\nu_4})
\nonumber\\
&\;\times\left\{[1-n(\epsilon_{\vec{k}_1\nu_1})][1-n(\epsilon_{\vec{k}_2\nu_2})]n(\epsilon_{\vec{k}_3\nu_3})n(\epsilon_{\vec{k}_4\nu_4})\right.
\nonumber\\
&\;\;\;\left.-n(\epsilon_{\vec{k}_1\nu_1})n(\epsilon_{\vec{k}_2\nu_2})[1-n(\epsilon_{\vec{k}_3\nu_3})][1-n(\epsilon_{\vec{k}_4\nu_4})]\right\},
\label{eq:appendix-A1}
\end{align}
where $n(\epsilon_{\vec{k}\nu})\equiv n_{\nu}(\epsilon_{\vec{k}\nu})$ is a short-hand symbol in this appendix. We may insert resolutions of unity
\begin{align}
\int dE\,\delta(E-\epsilon_{\vec{k}\nu})=1
\end{align}
for $\vec{k}_2,\vec{k}_3,\vec{k}_4$ on the right-hand side of Eq.~\eqref{eq:appendix-A1} to get
\begin{align}
&\!\!
\frac{dn(\epsilon_{\vec{k}_1\nu_1})}{dt}=\frac{2\pi}{\hbar}\frac{1}{N^2}\sum_{\vec{k}_2\vec{k}_3\vec{k}_4}\sum_{\nu_2\nu_3\nu_4}|\tilde{U}_{\nu_1\nu_2\nu_3\nu_4}(\vec{k}_1\vec{k}_2\vec{k}_3\vec{k}_4)|^2
\nonumber\\
&\;\;\times\delta_{\vec{k}_1+\vec{k}_2,\vec{k}_3+\vec{k}_4}\int dE_2dE_3dE_4
\delta(\epsilon_{\vec{k}_1\nu_1}\!\!+E_2-E_3-E_4)\nonumber\\
&\;\;\times\delta(E_2-\epsilon_{\vec{k}_2\nu_2})\delta(E_3-\epsilon_{\vec{k}_3\nu_3})\delta(E_4-\epsilon_{\vec{k}_4\nu_4})
\nonumber\\
&\;\;\times\left\{[1-n(\epsilon_{\vec{k}_1\nu_1})][1-n_{\nu_2}(E_2)]n_{\nu_3}(E_3)n_{\nu_4}(E_4)\right.
\nonumber\\
&\quad\;\left.-n(\epsilon_{\vec{k}_1\nu_1})n_{\nu_2}(E_2)[1-n_{\nu_3}(E_3)][1-n_{\nu_4}(E_4)]\right\}.\!
\label{eq:appendix-A3}
\end{align}
Multiplying by $\frac{1}{N}\delta(E_1-\epsilon_{\vec{k}_1\nu_1})$ on both sides of Eq.~\eqref{eq:appendix-A3} and summing over $\vec{k}_1$, the left-hand side becomes
\begin{align}
\mathrm{LHS}&=\frac{1}{N}\sum_{\vec{k}_1}\delta(E_1-\epsilon_{\vec{k}_1\nu_1})\frac{dn_{\nu_1}(E_1)}{dt}
\nonumber\\
&=\frac{d}{dt}D_{\nu_1}(E_1)n_{\nu_1}(E_1)=\frac{dN_{\nu_1}(E_1)}{dt},
\end{align}
using notations defined in Eqs.~\eqref{eq:density-of-states}--\eqref{eq:density-of-holes} of the main text. The right-hand side of Eq.~\eqref{eq:appendix-A3} becomes
\begin{align}
\mathrm{RHS}&=\frac{2\pi}{\hbar}\frac{1}{N^3}\!\sum_{\nu_2\nu_3\nu_4}\int dE_2dE_3dE_4\,\delta(E_1\!+\!E_2\!-\!E_3\!-\!E_4)
\nonumber\\
&\times\sum_{\{\vec{k}\}}|\tilde{U}_{\nu_1\nu_2\nu_3\nu_4}(\vec{k}_1\vec{k}_2\vec{k}_3\vec{k}_4)|^2\delta_{\vec{k}_1+\vec{k}_2,\vec{k}_3+\vec{k}_4}
\nonumber\\
&\times\delta(E_1-\epsilon_{\vec{k}_1\nu_1})\cdots\delta(E_4-\epsilon_{\vec{k}_4\nu_4})
\nonumber\\
&\times\left\{[1-n_{\nu_1}(E_1)][1-n_{\nu_2}(E_2)]n_{\nu_3}(E_3)n_{\nu_4}(E_4)\right.
\phantom{\frac{1}{2}}
\nonumber\\
&\left.-n_{\nu_1}(E_1)n_{\nu_2}(E_2)[1\!-\!n_{\nu_3}(E_3)][1\!-\!n_{\nu_4}(E_4)]\right\}\!.\!
\end{align}
Up to this point, the treatment is exact. Here comes the
\linebreak 
approximation: the matrix element modulus squared, $|\tilde{U}_{\nu_1\nu_2\nu_3\nu_4}(\vec{k}_1\vec{k}_2\vec{k}_3\vec{k}_4)|^2$, together with the Kronecker $\delta_{\vec{k}_1+\vec{k}_2,\vec{k}_3+\vec{k}_4}$ that can be thought of as already contained in $|\tilde{U}_{\nu_1\nu_2\nu_3\nu_4}(\vec{k}_1\vec{k}_2\vec{k}_3\vec{k}_4)|^2$, is replaced by the k-averaged quantity $\overline{|U|^2}_{\nu_1\nu_2\nu_3\nu_4}$ in Eq.~\eqref{eq:U2-kavg} times $1/N$. Then
\begin{align}
&\frac{dN_{\nu_1}(E_1)}{dt}=\frac{2\pi}{\hbar}\!\sum_{\nu_2\nu_3\nu_4}\int dE_2dE_3dE_4\,\delta(E_1\!+\!E_2\!-\!E_3\!-\!E_4)\nonumber\\
&\;\;\times\frac{1}{N^4}\sum_{\{\vec{k}\}}\overline{|U|^2}_{\nu_1\nu_2\nu_3\nu_4}\delta(E_1-\epsilon_{\vec{k}_1\nu_1})\cdots\delta(E_4-\epsilon_{\vec{k}_4\nu_4})
\nonumber\\
&\;\;\times\left\{[1-n_{\nu_1}(E_1)][1-n_{\nu_2}(E_2)]n_{\nu_3}(E_3)n_{\nu_4}(E_4)\right.
\nonumber\\
&\quad\left.-n_{\nu_1}(E_1)n_{\nu_2}(E_2)[1-n_{\nu_3}(E_3)][1-n_{\nu_4}(E_4)]\right\}.
\end{align}
Since $\overline{|U|^2}_{\nu_1\nu_2\nu_3\nu_4}$ is independent of $\{\vec{k}\}=\vec{k}_1\ldots\vec{k}_4$, it can be taken out of the summation over $\{\vec{k}\}$, which then gives us the product $D_{\nu_1}(E_1)D_{\nu_2}(E_2)D_{\nu_3}(E_3)D_{\nu_4}(E_4)$ of four densities of states. Then using notations in Eqs.~\eqref{eq:density-of-electrons}--\eqref{eq:density-of-holes} of the main text, we have
\begin{align}
&\frac{dN_{\nu_1}(E_1)}{dt}=\frac{2\pi}{\hbar}\sum_{\nu_2\nu_3\nu_4}\overline{|U|^2}_{\nu_1\nu_2\nu_3\nu_4}\int dE_2dE_3dE_4
\nonumber\\
&\;\;\times\delta(E_1+E_2-E_3-E_4)
\nonumber\\
&\;\;\times\left[\bar{N}_{\nu_1}(E_1)\bar{N}_{\nu_2}(E_2)N_{\nu_3}(E_3)N_{\nu_4}(E_4)\right.
\nonumber\\
&\quad\left.-N_{\nu_1}(E_1)N_{\nu_2}(E_2)\bar{N}_{\nu_3}(E_3)\bar{N}_{\nu_4}(E_4)\right],
\end{align}
which reproduces Eq.~\eqref{eq:QBE-kavg}. The main assumptions are the slow manifold assumption in Eq.~\eqref{eq:occ-num}, which reduces the number of dynamical degrees of freedom, and the local interaction and random band approximation, which justify the k-averaging of the rate constants.

\begin{figure}
\centering
\includegraphics[width=0.78\columnwidth]{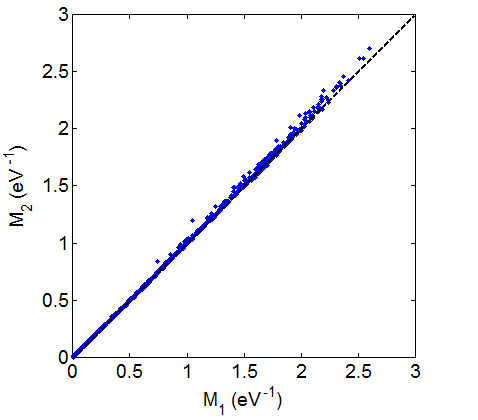}
\caption{(Color online) Integration measures $M_1$ and $M_2$ for different band indices $\nu_1,\nu_2,\nu_3,\nu_4$ in VO$_2$. The energy delta functions in Eqs.~\eqref{eq:appendix-A8}--\eqref{eq:appendix-A9} are smeared to a finite width of $\pm 5\!$~meV, which is compatible with the k-point mesh of $20\times 20\times 20$ we used. \label{fig:M1-M2}}
\end{figure}

In the actual implementation of Eq.~\eqref{eq:U2-kavg} to obtain the k-averaged rate constants $\overline{|U|^2}_{\nu_1\nu_2\nu_3\nu_4}$, it is more convenient to first randomly generate matrix elements $|\tilde{U}_{\nu_1\nu_2\nu_3\nu_4}(\vec{k}_1\vec{k}_2\vec{k}_3\vec{k}_4)|^2$ that satisfy both momentum and energy conservation, take the sample average and then multiply the result by a correction factor $M_2/M_1$, where the integration measure
\begin{align}
M_1=\frac{1}{N^4}\sum_{\{\vec{k}\}}\delta(\epsilon_{\vec{k}_1\nu_1}\!\!+\epsilon_{\vec{k}_2\nu_2}\!\!-\epsilon_{\vec{k}_3\nu_3}\!\!-\epsilon_{\vec{k}_4\nu_4}),
\label{eq:appendix-A8}
\end{align}
does not consider momentum conservation, while
\begin{align}
M_2=\frac{1}{N^3}\sum_{\{\vec{k}\}}\delta_{\vec{k}_1+\vec{k}_2,\vec{k}_3+\vec{k}_4}\delta(\epsilon_{\vec{k}_1\nu_1}\!\!+\epsilon_{\vec{k}_2\nu_2}\!\!-\epsilon_{\vec{k}_3\nu_3}\!\!-\epsilon_{\vec{k}_4\nu_4})
\label{eq:appendix-A9}
\end{align}
does. But it turns out that $M_1\approx M_2$ according to our actual data for VO$_2$ (see Fig.~\ref{fig:M1-M2}). Therefore the correction factor $M_2/M_1$ is insignificant and this also partly justifies the random band approximation: the fact that $\vec{k}_1+\vec{k}_2=\vec{k}_3+\vec{k}_4$ does not make $M_2$ too different from the case that $\vec{k}_1+\vec{k}_2$ equals any other value. The k-points are irrelevant and all that matter are the energies. Therefore whether momentum is conserved or not in doing the k-averaging makes no big difference.

\section{Absorption percentage of laser energy \label{sec:appendix-B}}

In this appendix, we present our calculation of the laser energy absorbed by the VO$_2$ sample in \cite{Morrison14}.

\begin{figure}[h!]
\centering
\includegraphics[width=0.75\columnwidth]{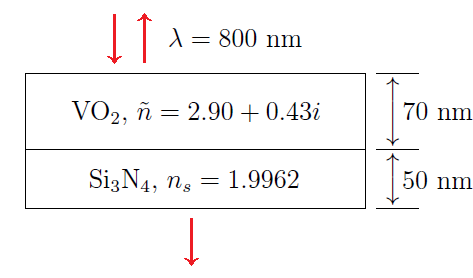}
\caption{(Color online) The setup of Morrison's pump-probe experiment of a VO$_2$ thin film on top of a Si$_3$N$_4$ substrate.
\label{fig:Exp-setup}}
\end{figure}

Ref.~\cite{Verleur68} gives the complex dielectric constant $\tilde{\epsilon}=8.2+2.5i$ of the laser wavelength $\lambda=800$~nm, which yields a complex index of refraction $\tilde{n}=\sqrt{\tilde{\epsilon}}=2.90+0.43i$. The index of refraction of the Si$_3$N$_4$ substrate $n_s=1.9962$ of $\lambda=800$~nm can also be found online. The thicknesses of the VO$_2$ sample and the Si$_3$N$_4$ substrate $d_1=70$~nm and $d_2=50$~nm are given in the supplementary material of \cite{Morrison14}. These data allow us to reconstruct the experimental setup in Fig.~\ref{fig:Exp-setup}.

Since the duration of the laser pulses used in \cite{Morrison14} is 35~fs, which is equivalent to over 13 oscillation periods of the 800~nm laser, the absorption of energy from the laser pulse can be obtained to adequate approximation by solving steady-state wave equations. Nonlinear optical effects are neglected \textit{a posteriori} because the density of excited particle-hole pairs is small. We may then use the formulas given in \cite{Tomlin68}, assuming normal incidence ($<10^\circ$ according to supplementary of \cite{Morrison14}).

The numerical result is that $R=43$~\% of the incident fluence gets reflected, $T=38$~\% gets transmitted, and the remaining $\Delta=1-R-T=19$~\% gets absorbed. Since the decay length of intensity in the VO$_2$ sample is $\lambda/4\pi\,\mathrm{Im}\,\tilde{n}=148$~nm$\,\gg d_1=70$~nm, the absorbed laser energy is roughly uniformly distributed  throughout the sample along the thickness direction.

Then using the density data $\rho=4.571$~g/cm$^3$ online for the $M_1$ phase of VO$_2$, we can find the unit cell volume and energy increase per unit cell under an incident laser fluence of $3.7\sim 9$~mJ/cm$^2$. The result equals $\Delta E_\mathrm{tot}=0.074\sim 0.18$~eV, the number used in the main text.

\bibliography{VO2_refs}
\end{document}